\documentclass[12pt]{iopart}
\usepackage{amssymb,graphicx}
\usepackage{color}
\usepackage[utf8]{inputenc}
\usepackage{hyperref}
\usepackage{caption}
\usepackage{subcaption}
\usepackage{url}
\usepackage{cite}
\usepackage{color}

\begin{document}

\title{Cosmic expansion from spinning black holes}

\author{John T. Giblin, Jr${}^{1,2}$}
\author{James B. Mertens${}^{3,4,5}$}
\author{Glenn D. Starkman${}^{2}$}
\author{Chi Tian${}^{2}$}
\ead{cxt282@case.edu}

\begin{abstract}
We examine how cosmological expansion arises in a universe containing a lattice of spinning black holes. We study averaged expansion properties as a function of fundamental properties of the black holes, including the bare mass of the black holes and black hole spin. We then explore how closely the expansion properties correspond to properties of a corresponding matter-dominated FLRW universe. As residual radiation present in the initial data decays, we find good agreement with a matter-dominated FLRW solution, and the effective density in the volume is well-described by the horizon mass of the black hole.
\end{abstract}


\address{${}^1$Department of Physics, Kenyon College, 201 N College Rd, Gambier, OH 43022}
\address{${}^2$CERCA/ISO, Department of Physics, Case Western Reserve University, 10900 Euclid Avenue, Cleveland, OH 44106}
\address{${}^3$Department of Physics and Astronomy, York University, Toronto, Ontario, M3J 1P3, Canada}
\address{${}^4$Perimeter Institute for Theoretical Physics, Waterloo, Ontario N2L 2Y5, Canada}
\address{${}^5$Canadian Institute for Theoretical Astrophysics, University of Toronto, Toronto, ON M5H 3H8 Canada}

\maketitle

\section{Introduction}

Cosmological systems are often modeled as perturbations around a
homogeneous, isotropic Friedmann-Lema\^{i}tre-Robertson-Walker (FLRW) spacetime where the background dynamics are described by homogeneous and isotropic stress-energy sources. 
Yet, the homogeneity of the stress-energy tensor is manifestly
broken on smaller length scales where discrete objects exist,
and inhomogeneous structures and well-isolated astrophysical systems predominate.
A notion of homogeneity and isotropy can still be recovered through averaging,
in which small-scale structures in the Universe are coarse-grained and an
effective FLRW cosmology emerges.
Observations support the picture that we live in an approximately homogeneous and isotropic FLRW universe\cite{Aghanim:2018eyx,Komatsu_2011};
what remains uncertain is the precise relationship between the actual Universe, with its extreme inhomogeneity on many scales, and the perturbed homogeneous isotropic cosmology we use as a model.

In recent years, numerical studies have begun to explore some of the differences
between these pictures\cite{Giblin:2015vwq,Bentivegna:2015flc}, giving rise to many questions including: how do beams of light
and gravitational waves propagate in a warped cosmological vacuum rather than
a perturbed perfect fluid\cite{0907.4109,1204.2411}?
What microphysics best describes the
manner in which isolated objects contribute to global cosmological expansion?

In this work, we begin to examine the latter of these questions by simulating a lattice of
spinning black holes and examining properties of the spacetime, which is shown to develop
an overall average FLRW-like cosmological expansion.
Black hole lattice models have been employed as toy models for cosmological systems in order to ask
such questions in the past \cite{0907.4109,1204.2411,1204.3568,1306.1389,1306.4055,1307.7673,1309.2876,1404.1435}
(and see \cite{1801.01083} for a recent review), and these models and similar
semianalytic models have been able to provide insights into both the physics of spatial
hypersurfaces in these models and, more recently, observables \cite{Bentivegna:2016fls,Sanghai:2017yyn,Fleury:2017owg}.
Such models have been found to reproduce FLRW-type behavior with varying degrees
of fidelity, with properties that depend on the precise details of the
inhomogeneous structure \cite{Korzynski:2013tea,Korzynski:2014nna,Fleury:2018cro,Durk:2017rky}. This dependence
is interesting in and of itself, as it suggests that the cosmological properties
of inhomogeneous spacetimes do not always provide insight into the more
fundamental small-scale properties of a spacetime. For example, different measures
of the mass contained within these spacetimes have been found to disagree by
orders of magnitude: some definitions of mass appear to coincide
with FLRW expectations, while others do not \cite{1203.6478,1204.3568,1405.3197}.

Here we extend these models to a lattice of spinning (rather than purely static) black holes.
We lay down initial conditions and follow the subsequent evolution of the spacetime
using numerical general-relativistic simulations. For black holes parametrized by a
given mass and spin, we examine the expansion rate within the box, and explore
how different energy components contribute to cosmological
expansion. Although we do not calculate cosmological observables here,
this work lays down the foundation for a series of future work regarding
observational consequences.

We first describe our procedure for setting initial conditions with a
spinning black hole in a periodic spacetime in Section~\ref{sec:ICs}.
This is motivated by previous solutions
found within the conformal-transverse-traceless decomposition \cite{1204.2411},
now extended to obtain a solution similar to the Bowen-York solution
in the vicinity of the black hole. We provide details of the numerical
scheme used to evolve the spacetime in Section~\ref{sec:evolution}, and review
definitions of mass useful for characterizing properties of the spacetime.
In Section~\ref{sec:results}
we describe the different contributions to the Hamiltonian constraint equation,
showing how various terms contribute to cosmological expansion.
We find that our initial conditions contain a substantial anisotropic energy
density exterior to the black hole that quickly decays; the remaining spinning
black hole sources curvature which continues to give rise to expansion within
the lattice. 
We then evaluate the behavior of different statistical measures of lattice
properties, concluding that volume-averaged properties appropriately describe the behavior,
and finding that the horizon mass, including both the irreducible (bare) mass and the
angular momentum, is sufficient for describing the observed expansion.
Lastly, we examine the averaged expansion rate, and compare this to the
cosmological expansion rate one might infer based on the mass of the black hole.
We find that the expansion rate initially behaves as a mixture of matter and radiative
content, consistent with residual radiation present in the initial data, with
the radiative content decaying and matter-dominated behavior emerging.

\section{Creating a spinning-black-hole lattice cosmology}
\label{sec:ICs}

We begin by reviewing the 3+1 decomposition of Einstein's equations, and
writing the constraint equations from this formalism in a form suitable
for numerically setting initial conditions with spinning black holes. We
restrict this discussion to vacuum solutions, although this formalism can
be generalized to include stress-energy sources. We will in particular make
use of the conformal transverse-traceless (CTT) decomposition of Einstein's
equations \cite{York:1978}, which extends the standard 3+1 decomposition,
in order to obtain solutions on spatial hypersurfaces.

We begin by writing the line element as
\begin{equation}
  ds^{2}=-\alpha^{2}dt^{2}+\gamma_{ij}\left(dx^{i}+\beta^{i}dt\right)\left(dx^{j}+\beta^{j}dt\right)\,.
\end{equation}
The non-dynamical Einstein's equations, projected onto spatial hypersurfaces
described by this metric, can be written as
\begin{eqnarray}
\label{eq:constraints}
  R + K^2 - K_{ij}K^{ij} &=& 0, \\
  D_j K^{j}_{\;i} - D_i K & = & 0,\nonumber
\end{eqnarray}
respectively known as the Hamiltonian and momentum constraint equations.
The derivatives $D_i$ are covariant with respect to the 3-metric $\gamma_{ij}$
  and $R$ is the associated 3-dimensional Ricci scalar.
The extrinsic curvature, $K_{ij}$, can be further decomposed into its trace, $K$,
and a traceless tensor, $A_{ij}$, 
\begin{eqnarray}
\label{eq:2}
K_{ij} = A_{ij} + \frac13 \gamma_{ij} K.
\end{eqnarray}
In an FLRW model, the trace, $K$, parameterizes the Hubble expansion rate with $H_{\rm FLRW} = -K/3$,
while $K$ is instead zero on time-symmetric hypersurfaces, for example
asymptotically flat spacetimes in appropriate coordinates\cite{Bowen:1980yu}.

The 3-metric can also be conformally decomposed,
$\gamma_{ij} = \Psi^4 \tilde{\gamma}_{ij}$, $A^{ij} = \Psi^{-10} \hat{A}^{ij}$,
allowing us to rewrite the constraint equations (\ref{eq:constraints}) in terms
of these new variables,
\begin{eqnarray}
\label{eq:basic_constraints}
  &\tilde{D}_i \tilde{D}^i \Psi - \frac{1}{8} \tilde{R} \Psi + \frac{1}{8} \hat{A}_{ij} \hat{A}^{ij} \Psi^{-7} + 2 \pi \Psi^5 - \frac{1}{12} K^2 \Psi^5 = 0 \\
  &\tilde{D}_j \hat{A}^{ij} - \frac{2}{3} \Psi^6 \tilde{D}^i K = 8 \pi \Psi^{10} S^i\,, \nonumber 
\end{eqnarray}
where $\tilde{D}_i$ and $\tilde{R}$ are now associated with the conformal metric
$\tilde{\gamma}_{ij}$.
The CTT decomposition
further breaks $\hat{A}^{ij} $ into longitudinal and transverse pieces,
\begin{eqnarray}
\label{eq:15}
\hat{A}^{ij} = \hat{A}^{ij}_{L} + \hat{A}^{ij}_{TT}.
\end{eqnarray}
Here $\hat{A}^{ij}_{TT}$ is transverse and traceless, satisfying $\tilde{D}_j\hat{A}^{ij}_{TT} = 0$.
The longitudinal piece $\hat{A}^{ij}_{L}$ can be written in terms of a vector $X^i$ as
\begin{eqnarray}
\label{eq:ALij}
\hat{A}^{ij}_{L}  = \tilde{D}^i X^j + \tilde{D}^j X^i - \frac{2}{3} \tilde{D}_k X^k \tilde{\gamma}^{ij} \equiv \left( \tilde{L} X \right)^{ij}.
\end{eqnarray}
The transverse-traceless component $\hat{A}^{ij}_{TT}$
contains information about transverse gravitational radiation, and can be set to zero
in order to minimize the gravitational radiation content of a solution. However,
this will not completely eliminate gravitational radiation, which can be sourced
nonlinearly, especially in a strong-gravity regime such as we are considering here. The
longitudinal component, on the other hand, contains information about the ``vector mode''
content of the spacetime, including frame-dragging and anisotropic effects. Generally,
vector modes are ignored in a cosmological setting, but their presence here will be
important for obtaining spinning-black-hole solutions.

Based on the above arguments, we further simplify the constraints by
setting $\hat{A}_{ij}^{TT} = 0$ , and choosing the metric to be 3-conformally flat,
$\tilde{\gamma}_{ij} = \delta_{ij}$. We then obtain
\begin{eqnarray}
\label{eq:simplifiedconstraints}
  &\nabla^2\Psi + \frac{1}{8} \left( \tilde{L} X \right)_{ij} \left( \tilde{L} X \right)^{ij} \Psi^{-7} - \frac{1}{12} K^2 \Psi^5 = 0 \nonumber \\
  &\nabla^2 X^{i} + \frac13 \partial^i \partial_j X^j- \frac{2}{3} \Psi^6 \partial^i K = 0,
\end{eqnarray}
where $\nabla^2$ is the Cartesian Laplacian and 
\begin{eqnarray}
\label{eq:6}
\left( \tilde{L} X \right)^{ij} = \partial^i X^j  + \partial^j X^{i} - \frac23 \delta^{ij} \partial_k X^k.
\end{eqnarray}

When $K = 0$, and in an asymptotically flat spacetime, a solution for $X^i$ known as the Bowen-York solution is given by
\begin{eqnarray}
\label{eq:approximateKerr}
X^i=\tilde{\epsilon}^{ijk}\frac{x_j J_k}{r^3},
\end{eqnarray}
where $x^i$ are the Cartesian coordinates, $r$ is the coordinate distance from the origin, and
$J^i$ is a vector satisfying $\tilde{D}_i J_j = 0$.
Here $\tilde{\epsilon}^{ijk} \equiv \sqrt{\tilde{\gamma}} \epsilon^{ijk}$ is the 3D Levi-Civita tensor associated with the conformal metric
$\tilde{\gamma}_{ij}$, so that $\tilde{D}_i \tilde{\epsilon}^{ijk}=0$.

Substituting this solution (\ref{eq:approximateKerr}) into expression (\ref{eq:ALij}), 
we obtain
\begin{eqnarray}
\label{eq:ALijBowenYork}
\hat{A}^{ij}_{L} = \left( \tilde{L} X \right)^{ij}=\frac{6}{r^3}x^{(i}\tilde{\epsilon}^{j)kl}J_k x_l / r^2.
\end{eqnarray}
This solution is commonly considered to contain a spinning black hole with spin $J_k$ 
\cite{Bowen:1980yu}.
The value of $\hat{A}^{ij}_{L}$ given by (\ref{eq:ALijBowenYork})
agrees with that of a
Kerr black hole at spatial infinity, implying this is true, however near the black hole this solution is not
equivalent to the Kerr metric. The Bowen-York solution has been found to contain some residual gravitational
radiation, and a maximum possible spin of $\|J\|=0.93$ \cite{gr-qc/9710096}.
We can nevertheless use this solution as inspiration for constructing initial conditions
in a cosmological setting, where the spacetime is no longer asymptotically flat.

Due to the discontinuity at the boundary, the Bowen-York solution (\ref{eq:approximateKerr}) is incompatible with periodic boundary conditions.
However, following a procedure similar to \cite{1204.2411, 1204.3568}, 
we can regularize the solution by multiplying parameters $M$ and $J^{i}$ in the metric fields by a {\it transition function} 
\begin{eqnarray}
\label{eq:transnfn}
W(r; \sigma, l) = \left\{
             \begin{array}{lr}
             0 &  0 \leq r < l\\
             ((r-l-\sigma)^6\sigma^{-6}-1)^6 & l\leq r < l+\sigma\\
             1 & l+\sigma \leq r 
             \end{array}                                                            \right.,
\end{eqnarray}
such that $W = 0$ at the origin, and transitions to $W = 1$ over some
distance scale $\sigma$ beginning at $r=l$. The vector $X^i$ is then regularized as 
\begin{eqnarray}
\label{eq:XiBYinPB}
X^i\approx\tilde{\epsilon}^{ijk}\frac{x_j J_k}{r^3} (1-W(r)).
\end{eqnarray}
We can also regularize the solution to the conformal factor $\Psi$ as
\begin{eqnarray}
\label{eq:PsiBYinPB}
\Psi\approx 1 + \frac{ M }{2r} (1-W(r)).
\end{eqnarray}
Eqs. \ref{eq:XiBYinPB} and \ref{eq:PsiBYinPB} can then be used as an initial guess for
solving the constraint equations.

\begin{figure}[htb]
  \centering
    \begin{subfigure}{0.47\textwidth}
      \includegraphics[width=0.99\textwidth]{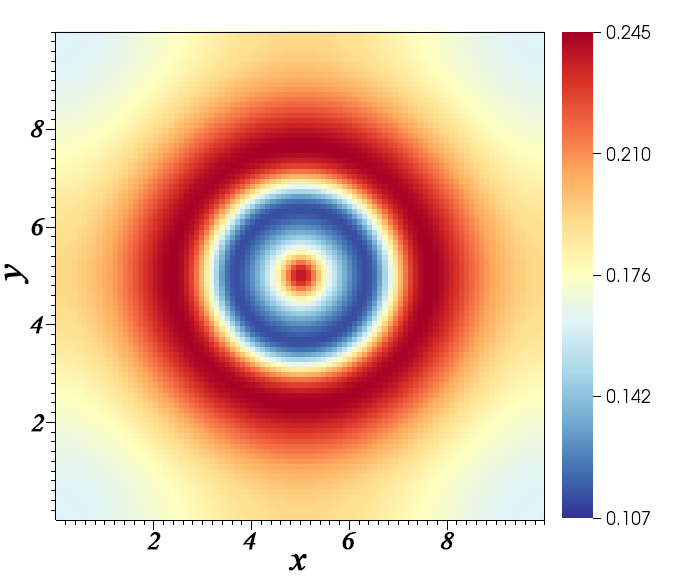}
    \end{subfigure}
    \begin{subfigure}{0.47\textwidth}
      \includegraphics[width=0.99\textwidth]{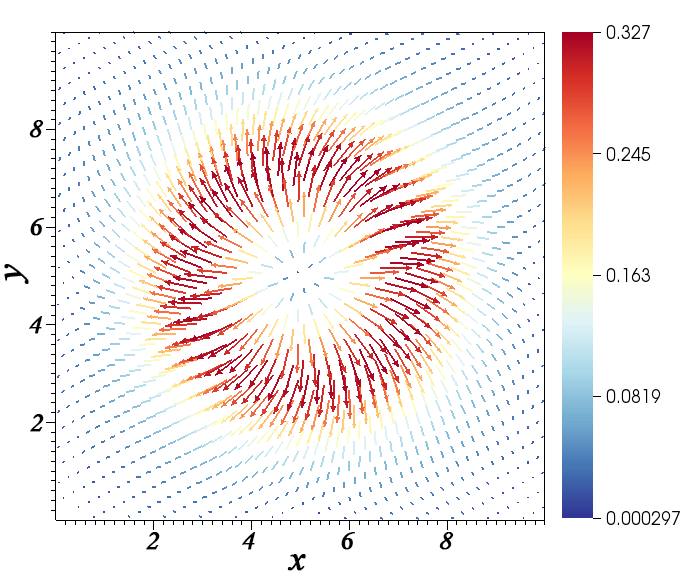}
    \end{subfigure}
    \caption{2D slices show the corrections to the initial guesses when solving Eq. \ref{eq:simplifiedconstraints} with a relaxation scheme.
      The initial guesses are given by Eqs. \ref{eq:XiBYinPB} and \ref{eq:PsiBYinPB} and the differences between those guesses and the
      exact solution are shown for field $\Psi$ (left) and $X^i$ (right).
      These also correspond to the fields $u$ and
      $X'^i$ as defined in Eq.~\ref{eq:16}. The choice of parameters for this solution correspond to R3 in Table \ref{t1}.
      } \label{init_u_X}
\end{figure}

In order for $K$ to be a negative constant at the boundaries, corresponding
to FLRW-like cosmological expansion, and zero in the center, corresponding to
a black hole solution, we also modulate the extrinsic curvature using $W(r)$,
\begin{eqnarray}
\label{eq:KinPB}
K = K_c W(r),
\end{eqnarray} 
where $K_c$ is a constant, similar to \cite{1306.1389}.

By plugging Eq. \ref{eq:KinPB} into the Hamiltonian and momentum
constraint equations and taking the approximate solution from Eqs.
\ref{eq:XiBYinPB} and \ref{eq:PsiBYinPB} to be an initial guess with $J_k = (0, 0, a)$
in Cartesian coordinates, we can proceed to solve for $\Psi$ and $X^i$.
The singularities in the solution are avoided by employing the so-called {\it puncture approach};
further details regarding this can be found in \ref{sec:init}.
In Fig. \ref{init_u_X} we show snapshots of the absolute difference (fields $u$ and $X'^i$
from Eqs. \ref{eq:16}) between the exact solution to Eqs. \ref{eq:simplifiedconstraints} and
the approximate solution of Eqs. Eq. \ref{eq:XiBYinPB} and \ref{eq:PsiBYinPB}. The main change
seen for $\Psi$, which is initially $\mathcal{O}(1)$ at the boundaries and much larger near the black
hole as per Eq.~\ref{eq:PsiBYinPB},
is an overall distortion of the physical volume of the spacetime, with additional radial corrections.
For $X^i$, the predominant correction is a large radial contribution in the transition region.

\section{Lattice Evolution}
\label{sec:evolution}

We solve the initial constraints and evolve the spacetime using
the grid-based numerical relativity code \textsc{CosmoGRaPH} \cite{Mertens:2015ttp}.
We first solve the constraint equations using an integrated elliptical-equation
solver, which employs a standard Full Multigrid (FMD) iteration scheme and an
inexact-Newton-relaxation method \cite{Press:2007:NRE:1403886}. We verify the resulting initial
conditions by checking that the Hamiltonian and momentum constraint equations are satisfied
with increasing precision as resolution is increased.

After setting initial conditions, spatial slices are advanced using the
BSSNOK formulation of numerical relativity \cite{Nakamura:1987zz,Shibata:1995we,Baumgarte:1998te},
with 4th order Runge-Kutta timestepping. All fields are discretized as
cell-centered data, and centered 4th-order finite-difference stencils are
used for all derivatives except for advection terms $\sim \beta^i\partial_i$, where
upwind derivatives are used instead. Note that the BSSNOK scheme evolves
$\tilde{A}_{ij} = \Psi^{-4} A_{ij}$,
rather than the $\hat{A}_{ij}$ defined when setting initial conditions.

The gauge condition used in our simulation is a revised version
of the widely employed ``1+log'' and ``Gamma-driver'' gauge condition:
\begin{eqnarray}
\label{eq:gaugecondition}
  \partial_t \alpha &=& -2 \eta \alpha (K - \langle K \rangle_{\rm edge}) + \beta^i \partial_i \alpha, \nonumber \\
  \partial_t \beta^i  & = & B^i,  \\
  \partial_t B^i & = & \frac34 \partial_t \Gamma^i - B^i. \nonumber
\end{eqnarray}
This differs from the usual ``1+log'' gauge by introducing a reference
expansion rate, $\langle K \rangle_{\rm edge}$, which is 
the conformal average of the extrinsic curvature $K$ along all edges of
the computational domain box defined in Eq. \ref{eq:21}.
This gauge choice has been demonstrated to have powerful singularity-avoidance properties\cite{Brown:2007pg};
the modification we make by subtracting $K$ relative
to the average boundary value allows the spatial slice to be driven
towards FLRW-like expansion away from the black hole.

Because of the collapsing nature of the black hole,
we have also integrated an Adaptive Mesh Refinement (AMR) framework
into the time evolution,
provided by the code SAMRAI \cite{DBLP:conf/sc/WissinkHKSE01},
an open-source structured adaptive-mesh-refinement application infrastructure.
By building hierarchies of grid levels with different resolution, dividing and
distributing patches into computational nodes, SAMRAI realizes high-efficiency
adaptive-mesh refinement and parallelization. To synchronize data on different levels,
we use tri-cubic Hermite interpolation\cite{kreyszig2019advanced}, which we find results in a high
degree of numerical stability.

We also require a technique to locate apparent horizons. Because of the aspherical
nature of spinning black holes, there is no symmetry of the apparent horizon
that makes it simple to locate. We therefore use the AHFinderDirect package
\cite{Thornburg:2003sf} to find the apparent horizon on a given spatial hypersurface.
The definition and the method of extracting angular momentum from
an isolated horizon come from
\cite{Dreyer:2002mx}, where the angular momentum of a black hole $J$ is defined as
\begin{eqnarray}
\label{eq:19}
J = \frac{1}{8\pi} \oint (\varphi^a R^b K_{ab}) \mathrm{d}^2V.
\end{eqnarray}
Note here $R^b$ is an outgoing vector normal to the horizon and
$\varphi^a$ is not a Killing vector of the full spacetime but a symmetry
vector defined locally on the horizon that preserves the induced metric $q_{ab}$,
so that
\begin{eqnarray}
\label{eq:induced}
 \mathcal{L}_{\varphi} q_{ab} = 0
\end{eqnarray}
(see \cite{Dreyer:2002mx} for more detail). The eigenvalue closest to unity
associated with the symmetry vector is within a percent of unity, indicating
the spacetime is very close to axisymmetric in the vicinity of the black hole.

We then track the black hole's irreducible mass, spin, and horizon mass, as well as
the expansion history of the spacetime.
The irreducible and horizon masses are defined as
\begin{eqnarray}
\label{eq:24}
M_{\rm H}^2 & \equiv & M_{\rm irr}^2 + \frac{J^2}{4M_{\rm irr}^2} \nonumber \\
M_{\rm irr}^2 & \equiv & \mathcal{A}/16\pi\,.
\end{eqnarray}
Here $\mathcal A$ is the area of the horizon, defined as
$\mathcal{A} \equiv \oint \sqrt{q} \mathrm{d}^2V$, where $q$ is the determinant
of the induced metric on the horizon.

In order to examine how well the spinning-black-hole-lattice universe
corresponds to a FLRW universe with similar expansion properties, or to
check how well the lattice obeys a Friedmann-like equation, we need to
define an effective density of the spacetime, $\rho_{\rm eff}$, and an average spacetime
expansion rate, $\left<K\right>$.
We then define a dimensionless parameter 
\begin{equation}
\label{eq:FLRW}
\mathcal{C} \equiv \frac{\rho_{\rm eff} }{\langle K \rangle^2 / 24\pi}\,. 
\end{equation}
According to the Friedmann equation, one should have $\mathcal{C}=1$
for an appropriately chosen ${\rho_{\rm eff} }$
and $\langle K \rangle$ as the effective Hubble parameter equals $\langle K \rangle /3$.

We first consider whether, in defining ${\rho_{\rm eff} }$
and $\langle K \rangle$, it is more appropriate to use a volume-averaging operation or to 
average over edges of the box.
We can set $\rho_{\rm eff} = \rho_{\rm edge,\,eff}$, given by
\begin{equation}
\label{eq:23}
\rho_{\rm edge,\,eff} \equiv M_{\rm eff} / (D_{\bot}^2 D_{\parallel}).
\end{equation}
$M_{\rm eff}$ can be either $M_{\rm H}$ or $M_{\rm irr}$, while
$D_{\bot}$ and $D_{\parallel}$ are distances along edges of the box in directions that
are perpendicular and  parallel to the spin direction respectively,
$D_i \equiv \int \mathrm{d}x^i \sqrt{\gamma_{ii}}$ for a Cartesian direction $i$
(where no sum over $i$ is implied).

An alternative is to make $\rho_{\rm eff} = \rho_{\rm vol,\,eff}$, with
\begin{equation}
\label{eq:rhovol}
\rho_{\rm vol,\,eff} \equiv M_{\rm eff} / V,
\end{equation}
where 
\begin{equation}
\label{eq:vol_defn}
V \equiv \int_{r>r_H} \mathrm{d}^3 x \sqrt{\gamma}
\end{equation}
is the conformal volume exterior to the black-hole horizon.

More generically, 
we define an averaged physical quantity $Q$ on the edge or volume 
\begin{eqnarray}
\label{eq:21}
  \left< Q\right>_{\rm edge} &\equiv&  \frac{\sum_{\rm all\;edges} \int \mathrm{d}x^i \sqrt{\gamma_{ii}} Q }{\sum_{\rm all\;edges} D_i} \\ \nonumber
  \left< Q \right>_{\rm vol} &\equiv& \frac{ 1 }{V} \int_{r>r_H} \mathrm{d}^3 x \sqrt{\gamma} Q\,.
\end{eqnarray}
We can now define the ratio of the left and right hand sides of the effective Friedmann equations as
\begin{eqnarray}
\label{eq:20}
  \mathcal{C}_{\rm edge,\,eff} = \frac{\rho_{\rm edge,\,eff}}{\langle K^2 \rangle_{\rm edge}  / 24 \pi}, \;\;\;\;\;\; \mathcal{C}_{\rm vol,\,eff} = \frac{\rho_{\rm vol,\,eff}}{\langle K^2 \rangle_{\rm vol}  / 24 \pi}.
\end{eqnarray}

\section{Results}
\label{sec:results}

In this section, we will present our main result. We will mainly focus on the
the expansion history and effects of spins. 
We will also investigate time evolution of dimensionless parameter $\mathcal{C}$ and evaluate
the effect of statistics.
\subsection{Initial condition effects on physical lattice properties}

The free parameters in our setup are the box size $L$, the mass scale $M$, the spin $a$,
the extrinsic curvature at the periodic boundary $K_c$,
and the parameters appearing in the transition function (\ref{eq:transnfn}), $l$, and $\sigma$.
To demonstrate the impact that varying these parameters has on the physical
properties of the spacetime (namely the black hole masses, densities, and $\mathcal{C}$),
we have listed these properties and  the corresponding
parameters in Table \ref{t1} for ten representative simulations.

\begin{table}[htbp]
  \caption{\label{t1} Parameters of initial setups}
  \begin{center}
  \item[]\begin{tabular}{l|ccccc|ccccc|c}
    \mr
    Runs &$L$& $a$& $K_c$ &$l$&$\sigma$& $M_{\rm irr}$ &$M_{\rm H}$       & $M_{\rm H}/V$&$\mathcal{C}_{{\rm edge},\,H}$ & $\mathcal{C}_{{\rm vol},\,H}$ & $c$\\
    \mr
    R1   &10  & 0  & -0.21 & 1 & 3.5    & 1.023    & 1.023  & 0.000414&0.897& 0.829 & 7.7\\
    R2   &10  & 0.6& -0.21 & 1 & 3.5    & 1.107    & 1.140  & 0.000404&0.894& 0.817 & 7.7\\
    R3   &10  & 0.9& -0.21 & 1 & 3.5    & 1.180    & 1.240  & 0.000395&0.891& 0.807 & 7.4\\
    R4   &10  & 0  & -0.21 & 0.1 & 3    & 0.982    & 0.982  & 0.000467&0.996& 0.844 & 8.3\\
    R5   &10  & 0.6& -0.21 & 0.1 & 3    & 1.072    & 1.108  & 0.000456&0.994& 0.828 & 7.4\\
    R6   &10  & 0.9& -0.21 & 0.1 & 3    & 1.147    & 1.212  & 0.000446&0.993& 0.815 & 8.1\\ 
    R7   &10  & 0.6& -0.15 & 1 & 3.5    & 1.246    & 1.269  & 0.000216&0.900& 0.842 & 7.6\\
    R8   &10  & 0.6& -0.1  & 1 & 3.5    & 1.453    & 1.468  & 0.000101&0.904& 0.865 & 7.6\\
    R9   &11  & 0.6& -0.21 & 1 & 3.5    & 1.058    & 1.096  & 0.000420&0.921& 0.827 & 7.4\\  
    R10  &12  & 0.6& -0.21 & 1 & 3.5    & 1.018    & 1.060  & 0.000432&0.941& 0.832 & 7.4
  \end{tabular}       
  \end{center}
\end{table}
In this table, only the first 2nd-6th columns are free parameters that were
chosen initially, while the 7th-11th columns are derived parameters that can only be calculated
after initial constraints are fully solved. The $\rho_{\rm eff}$ used 
in columns for $\mathcal{C}_{{\rm edge},\,H}$ and $\mathcal{C}_{{\rm vol},\,H}$ are calculated using $M_{\rm H}$.
The convergence rate for each run on initial slice is represented by parameter $c$ whose definition can be found in Eq. \ref{eq:c}.
In all runs, we do not vary $M$, instead choosing to work in units where $M=1$.

Examining these initial configurations, we can observe the following:
\begin{itemize}
\item Although the input parameter $M$ is equal to $M_{\rm H}$ for a Kerr spacetime
  with asymptotic flat boundary, the resulting $M_{\rm H}$ in the table only roughly tracks $M$,
  depending on other parameters as well.
\item Both  $\mathcal{C}_{{\rm edge},\,H}$ 
  and $\mathcal{C}_{{\rm vol},\,H}$ are somewhat less than
  1 initially, and change very little when
  the spin parameter $a$, box size $L$, or boundary extrinsic curvature $K_c$ are varied. This
  implies that the initial spatial slice is always ``close'' to FLRW.
\item Only by changing the combination of $l$ and $\sigma$ does $\mathcal{C}_{{\rm edge},\,H}$
  change significantly; However, the value of $\mathcal{C}_{{\rm vol},\,H}$ still does
  not change.
\item Increasing the box size $L$ (comparing R9 and R10 to R2)  does not increase the physical size of the box.
\item Changing the boundary extrinsic curvature $K_c$ (comparing R7 and R8 to R2) will
  change the effective density and physical box length significantly, but still keeps the ratio in the last
  two columns unchanged.
\end{itemize}

To summarize, the parameters that predominantly determine physical properties of the system are $K_c$, $M$, and $a$. These
strongly affect the simulation volume and black hole mass and spin. $L$, $l$, and $\sigma$
instead affect the coordinate description of the spacetime, and only weakly affect physical properties.

The initial value of the ratio $\mathcal{C}$ quantifies the deviation from the FLRW universe,
and is found to be relatively independent of our parameter choices.
We now wish to study its time evolution as well as the best way to fit
$\rho_{\rm eff}$ to a FLRW universe. Those topics are our main interests and will be
discussed in the following sections.

\subsection{Expansion properties and energy content}

We can now analyze the different contributions to the Hamiltonian constraint
equation, or the different ``energy'' contributions in the spacetime contributing
to expansion. Because we work in a vacuum spacetime, there is  no actual stress-energy
contribution, and all expansion must be a result of either curvature or kinetic terms
in the constraint equations. We can decompose these terms as in Hamilotnian constraint in Eq.~\ref{eq:basic_constraints} to get
\begin{eqnarray}
\label{modified_constraint}
 R / 8  +  \tilde{A}_{ij} \tilde{A}^{ij} / 8 +  K^2 / 12 = 0, 
\end{eqnarray}
and analyze the average behavior of: the curvature $\langle R \rangle / 8$; the anisotropic expansion term
$\langle \tilde{A}_{ij} \tilde{A}^{ij}\rangle / 8$, which contains contributions from
vector and tensor modes and their interactions; and the expansion itself,
$\left< K^2\right> / 12$.

\begin{figure}[htbp]
  \centering
  \begin{subfigure}{0.3\textwidth}
      \includegraphics[width=0.99\textwidth]{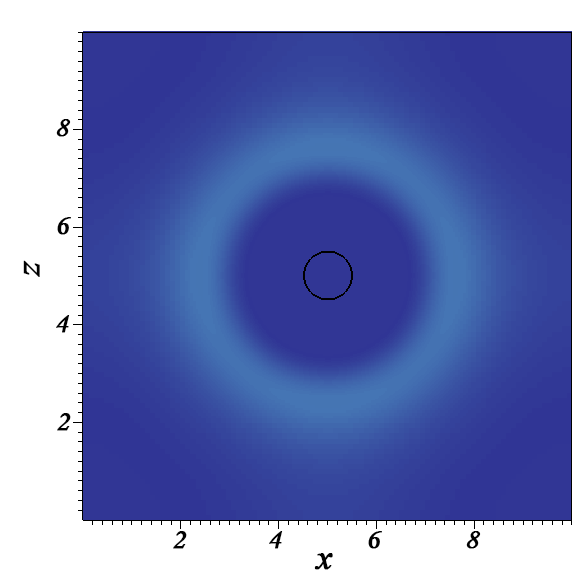}
    \end{subfigure}
    \begin{subfigure}{0.3\textwidth}
      \includegraphics[width=0.99\textwidth]{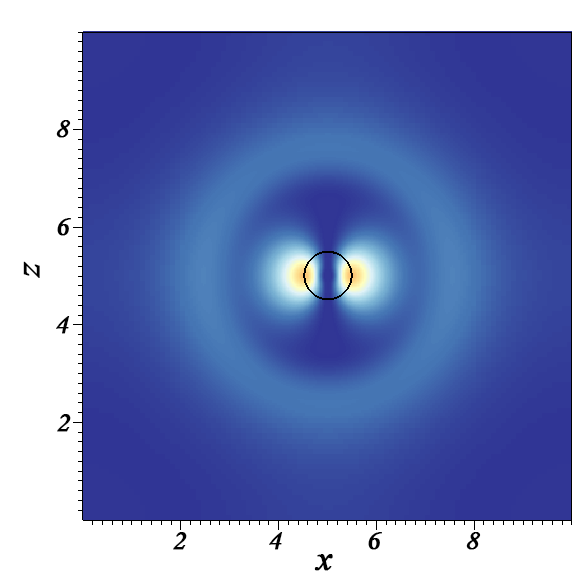}
    \end{subfigure}
    \begin{subfigure}{0.37\textwidth}
      \includegraphics[width=0.99\textwidth]{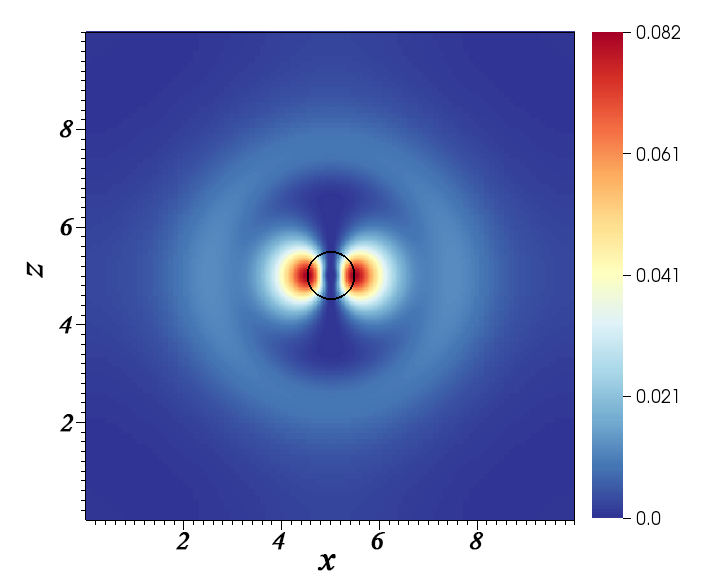}
    \end{subfigure}
    \begin{subfigure}{0.3\textwidth}
      \includegraphics[width=0.99\textwidth]{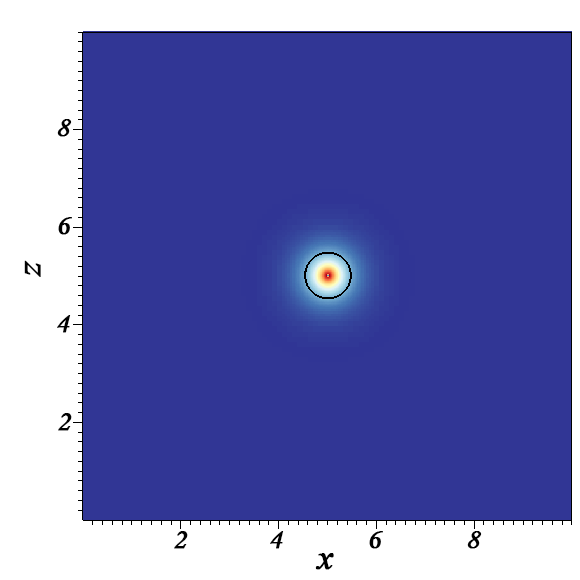}
    \end{subfigure}
    \begin{subfigure}{0.3\textwidth}
      \includegraphics[width=0.99\textwidth]{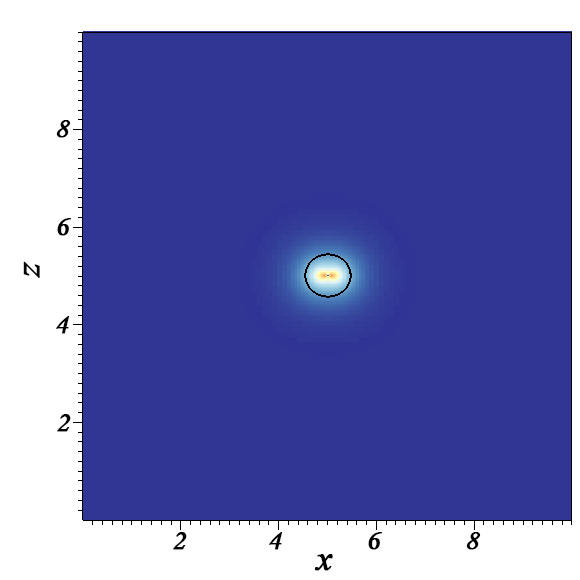}
    \end{subfigure}
    \begin{subfigure}{0.37\textwidth}
      \includegraphics[width=0.99\textwidth]{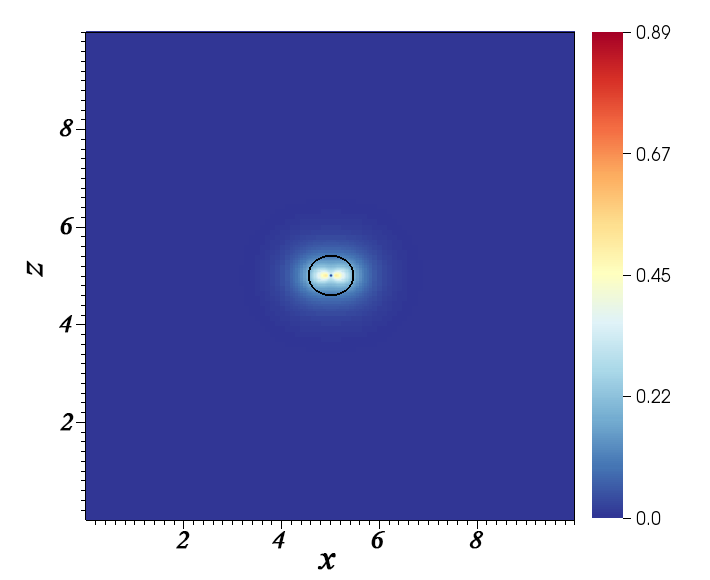}
    \end{subfigure}

    \caption{The conformally related trace-free part of the extrinsic curvature,
      $\tilde{A}_{ij}\tilde{A}^{ij}$, on two-dimensional slices that intersect
      the black hole through spatial hypersurfaces. The black circles are
      corresponding slices of apparent horizons.
      $\tilde{A}_{ij}\tilde{A}^{ij}$ is shown
      runs R1, R2 and R3 with spins $a = 0, 0.6, 0.9$ (left to right), and at times $t=0, 10$  (top to bottom).
      All quantities are in units where $M=1$.
    }
  \label{AijAij}
\end{figure}

In Fig.~\ref{AijAij}, we examine the contribution of the $\tilde{A}_{ij}\tilde{A}^{ij}$
term, which contains information about gravitational-wave and vector-mode energy content.
It shows that the vector and tensor modes are concentrated
near the black hole horizon, especially as the spacetime evolves and relaxes away from the
na\"ive initial conditions that we set. This is unsurprising in that vector and tensor modes
can be sourced nonlinearly in a strong-gravity regime, in contrast to the linear regime in a
cosmological setting, where they are expected to be negligible.
The absence of this contribution further away from the black hole shows that scalar
curvature is the dominant contribution to cosmological expansion.

\begin{figure}[htbp]
  \centering
      \begin{subfigure}{0.3\textwidth}
      \includegraphics[width=0.99\textwidth]{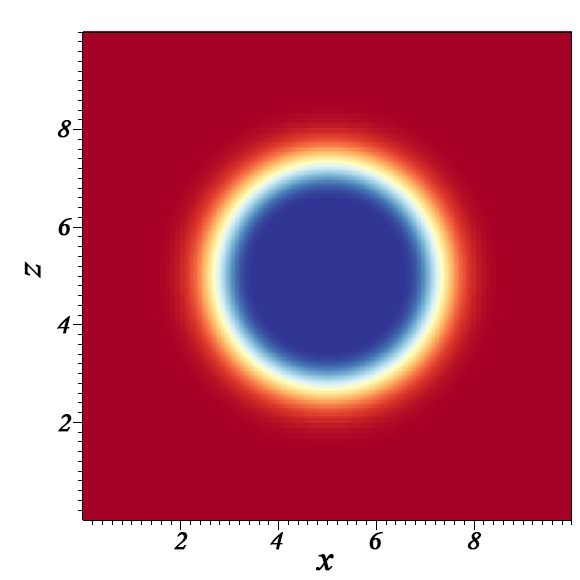}
    \end{subfigure}
        \begin{subfigure}{0.3\textwidth}
      \includegraphics[width=0.99\textwidth]{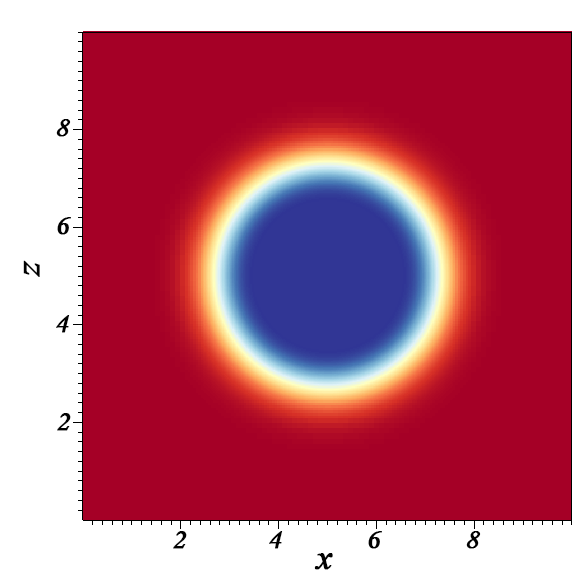}
    \end{subfigure}
    \begin{subfigure}{0.37\textwidth}
      \includegraphics[width=0.99\textwidth]{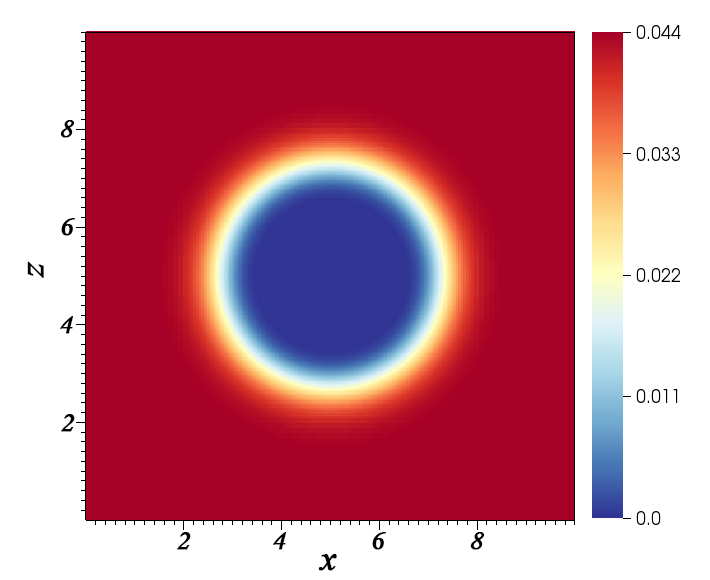}
    \end{subfigure}
    \begin{subfigure}{0.3\textwidth}
      \includegraphics[width=0.99\textwidth]{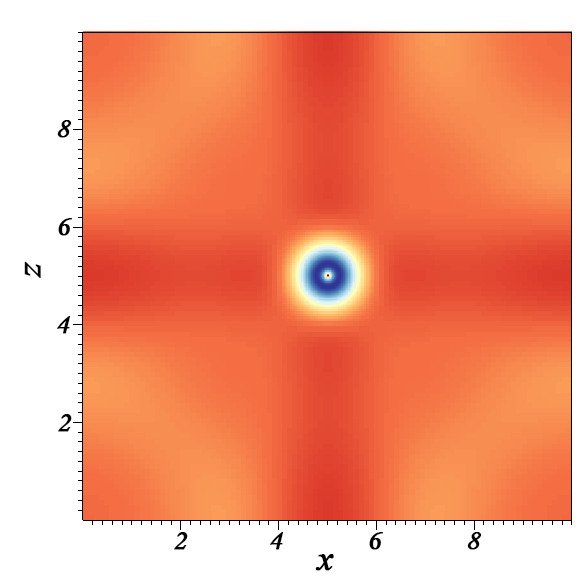}
    \end{subfigure}
        \begin{subfigure}{0.3\textwidth}
      \includegraphics[width=0.99\textwidth]{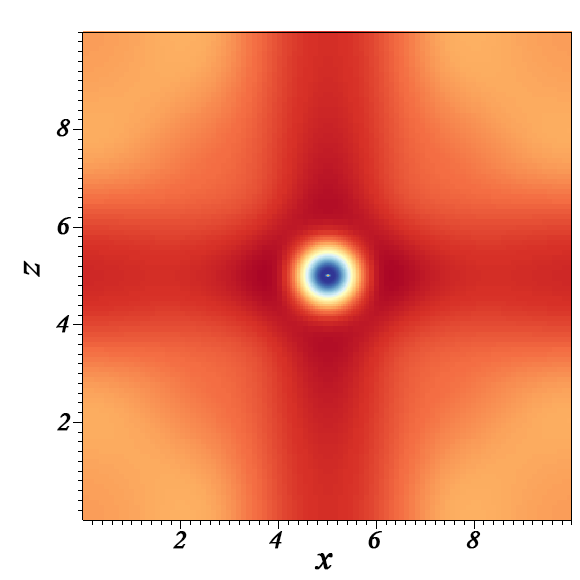}
    \end{subfigure}
    \begin{subfigure}{0.37\textwidth}
      \includegraphics[width=0.99\textwidth]{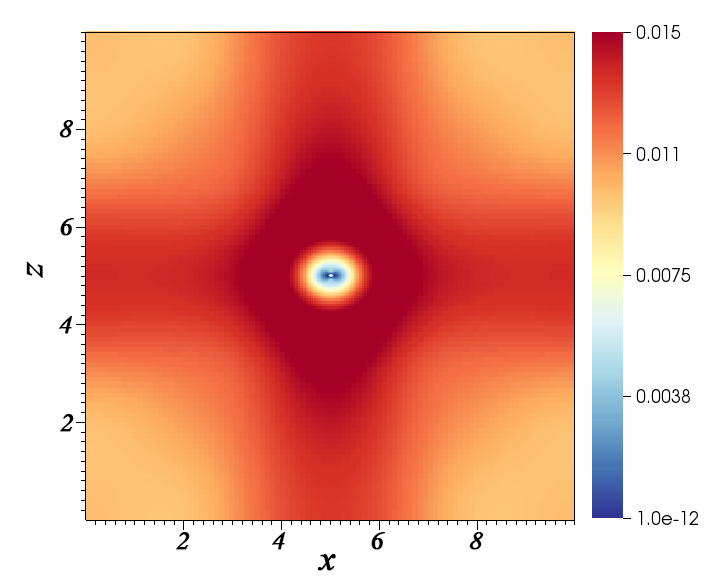}
    \end{subfigure}
    \begin{subfigure}{0.3\textwidth}
      \includegraphics[width=0.99\textwidth]{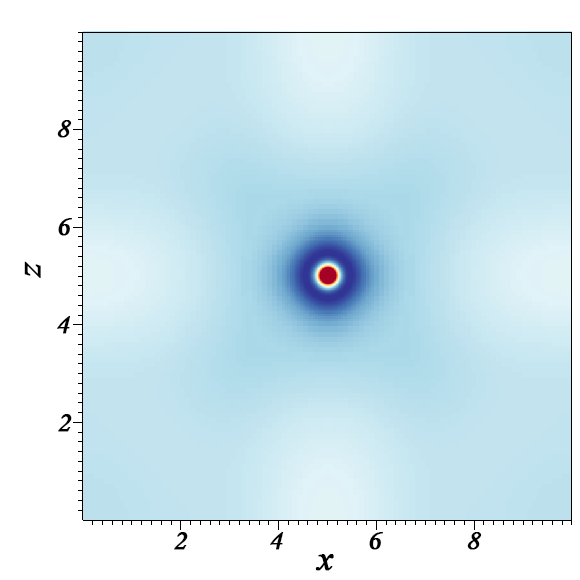}
    \end{subfigure}
        \begin{subfigure}{0.3\textwidth}
      \includegraphics[width=0.99\textwidth]{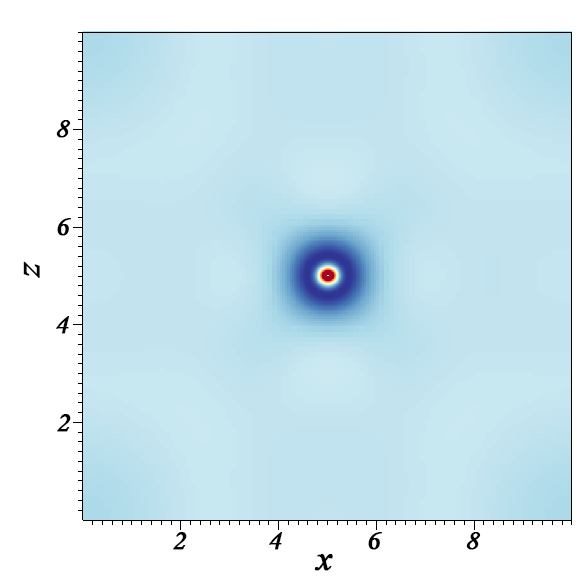}
    \end{subfigure}
    \begin{subfigure}{0.37\textwidth}
      \includegraphics[width=0.99\textwidth]{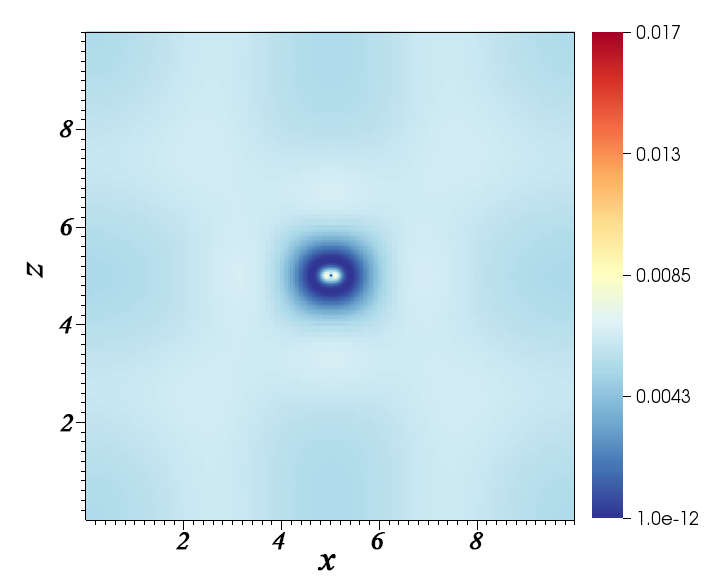}
    \end{subfigure}
    \caption{Time evolution snapshots of $K^2$ in R1, R2 and R3 with
$a = 0, 0.6, 0.9$ (from left to right), and $t=0 (M), 10 (M), 15(M)$ 
(from top to bottom).} 
  \label{Ksq}
\end{figure}

Fig. \ref{Ksq} similarly depicts the time evolution of $K^2$, which is the volume expansion
rate. A behavior very similar to $\tilde{A}_{ij} \tilde{A}^{ij}$ is identified --
deviations from cosmological-type expansion are found near the black hole, that gradually
become smooth far away from the black hole, especially as the simulation progresses.

Residual oscillations can be seen in the expansion rate, both spatially varying, and
as a function of time. The behavior of these oscillations depends on both the initial 
conditions and the gauge we choose, and we therefore do not consider these to be indicative of
an expansion rate that is physically oscillatory, ie. that would strongly impact the way
a geodesic observer would view the spacetime. We leave this speculation to future work,
although see also \cite{1801.01083}.

\begin{figure}[htbp]
  \centering
    \includegraphics[width=0.99\textwidth]{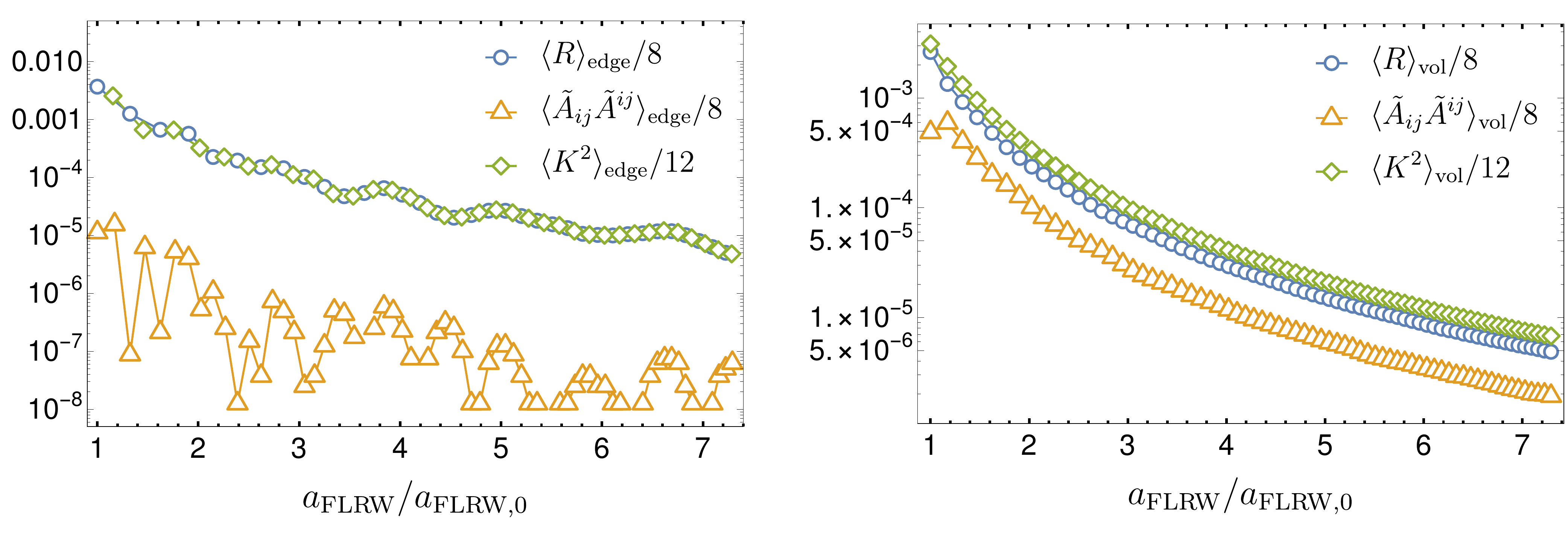}
  \caption{Behavior of the contributions $\left<R\right> / 8$, $\langle \tilde{A}_{ij}\tilde{A}^{ij}\rangle/8$ and $\left<K^2\right>/12$ to the Hamiltonian constraint in Eq. \ref{modified_constraint} in R2. The left panel shows edge-averaged terms, and the right panel shows results with volume-averaged terms. }
  \label{Aij_avg}
\end{figure}

We examine the different contributions to the Hamiltonian constraint equation more
quantitatively for R2 in Fig. \ref{Aij_avg}. This demonstrates that the vector and tensor
contributions $\langle \tilde{A}_{ij}\tilde{A}^{ij}\rangle$
are relatively small near the edge, 
but are appreciable when averaged over the volume exterior to the horizon.
It is important to note that this interpretation will be affected by gauge choice:
for example, the first-order gauge-invariant vector mode usually considered in a cosmological
setting, as well as true observables (eg. properties integrated along geodesics according
to observers), will contain a contribution from the shift that is not shown here.

\subsection{Expansion-mass correspondence}

As we evolve the spacetime using the gauge choice of Eq.~\ref{eq:gaugecondition},
lengths of edges as well as volume of
the spatial slice expand in the intuitively expected manner.
The coordinate
size of the black hole apparent horizon initially expands as the solution stabilizes, then
shrinks due to cosmological expansion, while the area stays the same.
We run the code until the black-hole horizon
becomes too small to be resolved accurately. At this moment, lengths in spatial slices
have roughly grown by a factor of $e^2$.

Because of the asymmetric setup, one might expect to see a different expansion rate in
different directions.
However, we find less than a $0.1\%$ difference in lengths along different edges of
the computational box ($D_{\parallel}$ and $D_{\bot}$),
and thus will ignore this discrepancy and focus on the quantities averaging on both 
parallel and perpendicular edges (see Eq. \ref{eq:21}).

The masses $M_{\rm irr}$ and $M_{\rm H}$ on the initial slice of each runs are shown in Table \ref{t1}.
During the evolution, their time dependence is nearly negligible: the
relative fluctuation in their values is as small as $0.1\%$ and dominated by
numerical uncertainty (see \ref{sec:convergence} for more detail),
consistent with the area theorem and conservation of angular momentum.

\begin{figure}[htbp]
  \centering
    \includegraphics[width=0.99\textwidth]{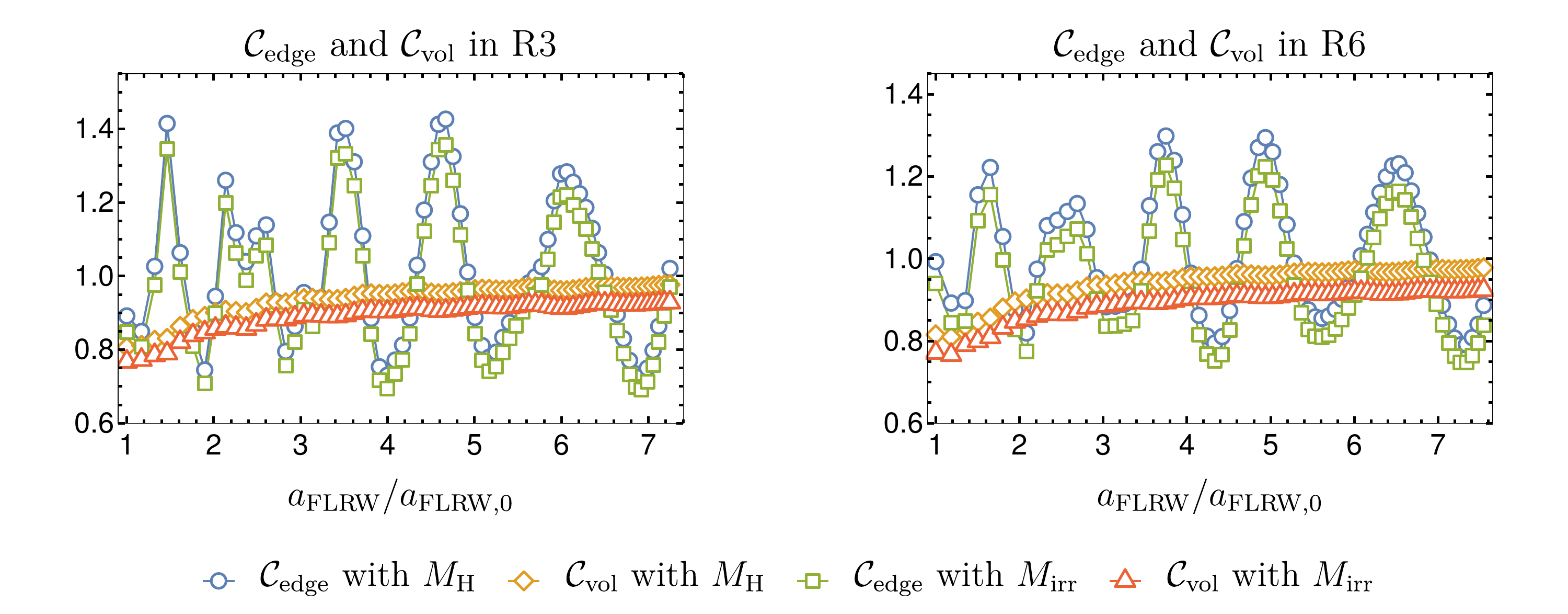}
    \caption{$\mathcal{C} = \rho_{\rm eff} / (\left<K\right>^2/24\pi)$ for runs R3 and R6.
      $\mathcal{C}$ is evaluated using edge and
      volume averages, with both $M_{\rm irr}$ and $M_{\rm H}$ used in $\rho_{\rm eff}$.
    }
  \label{ratios_M_H}
\end{figure}
\begin{figure}[htbp]
  \centering
    \includegraphics[width=0.99\textwidth]{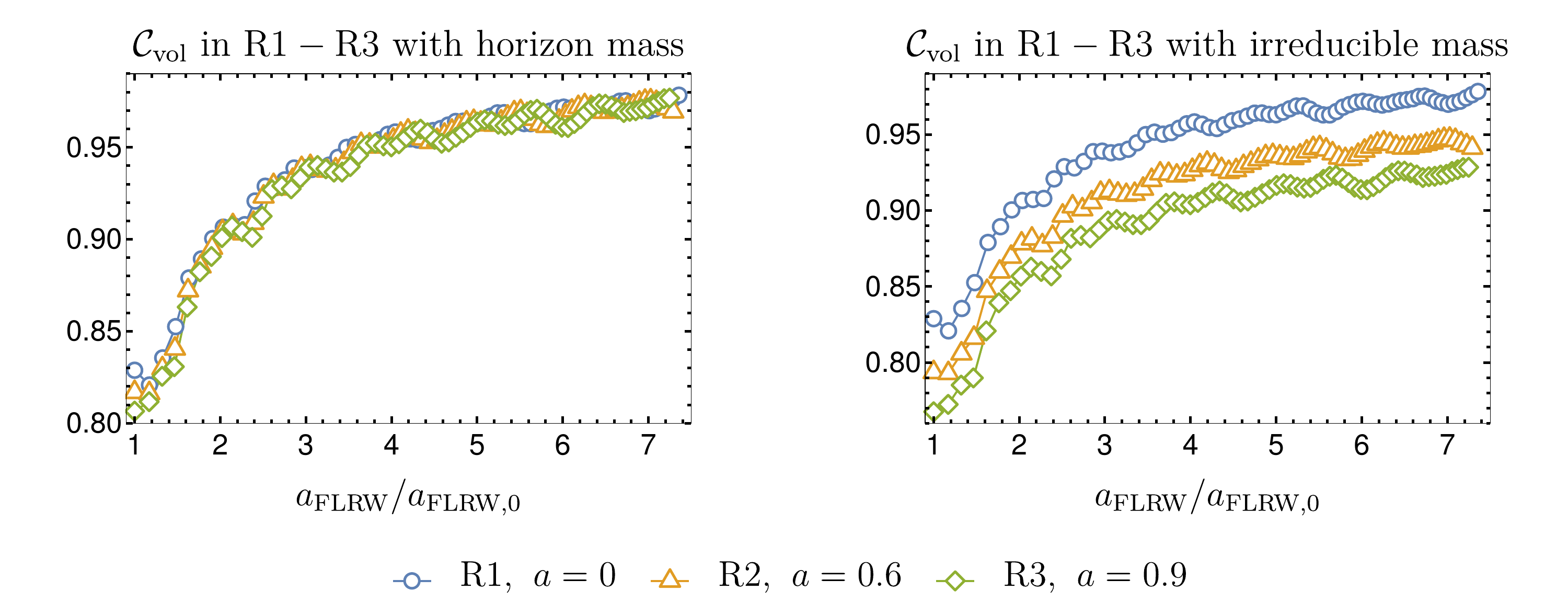}
    \caption{Behavior of the ratio $\mathcal{C}_{\rm vol}$ with different spins
    and masses. The horizon mass is used in the left panel, while the irreducible mass
    is used in the right.}
  \label{ratios_spins}
\end{figure}

From Fig. \ref{ratios_M_H}, we can see that 
$\mathcal{C}_{\rm edge}$ fluctuates with a large amplitude, consistent with
behavior seen in \cite{1306.4055}, while $\mathcal{C}_{\rm vol}$ gently increases to unity.
The difference between choices of irreducible mass and horizon mass results in a constant
shift between curves, which we investigate below.

We can attempt to reduce the amplitude of oscillations in the edge-averaged case
by adjusting $l$ and $\sigma$ to obtain  $\mathcal{C}_{\rm edge}\simeq 1$ on the initial slice. 
Comparing the panels in Fig.~\ref{ratios_M_H}, we see that this does not help in eliminating
the fluctuations.  The large amplitude fluctuations we see in $\mathcal{C}_{\rm edge}$ apparently arise from a combination of the way we slice the initial spacetime and the gauge condition, indicating
volume-averaged quantities appear to be a more appropriate representation of the physical behavior
of the system.

Fig. \ref{ratios_spins} provides us with more insight into
the spin dependence of $\mathcal{C}_{\rm vol}$ for different definitions of mass. 
$\mathcal{C}_{\rm vol}$ is approximately spin-independent when the horizon mass is used to construct $\rho_{\rm eff}$. It appears to approach the expected matter-dominated FLRW value of unity as the simulation evolves.  
Neither of these features are maintained when choosing $\rho_{\rm eff} = M_{\rm irr} / V$ (right panel).
The diminishing value of $\mathcal{C}_{\rm vol}$
as the spin is increased implies that an energy contribution to the Hamiltonian
constraint is not being accounted for.
The horizon mass $M_{\rm H}$ is thus a better
choice than $M_{\rm irr}$ when considering the effective mass in a spinning-black-hole-lattice universe.

\begin{figure}[tbp]
  \centering
    \includegraphics[width=0.99\textwidth]{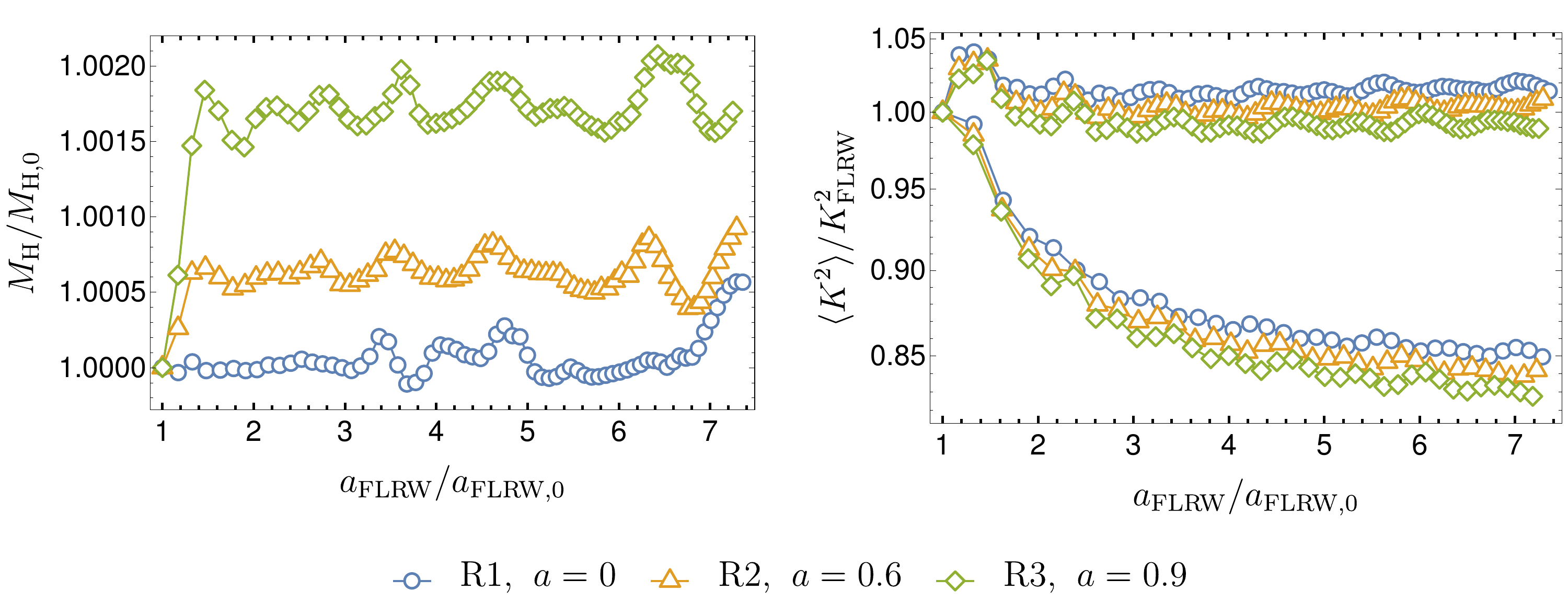}
    \caption{Dependence of $M_{\rm H} / M_{\rm H, 0}$ 
      and $\langle K^2 \rangle_{\rm vol}/\langle K^2 \rangle_{\rm vol, 0}$ as a function of scale factor.
      We also compare the averaged expansion rate to an FLRW model, according to a best fit function with radiation (top 3 lines) and without (bottom 3 lines), showing the actual expansion is well-described by a mix of matter and radiation.
    } 
  \label{rho_K_vs_a}
\end{figure}

Lastly, we consider how the averaged values we consider here map to corresponding
FLRW spacetimes in Fig.~\ref{rho_K_vs_a}. As a function of
FLRW scale factor, $a \equiv V^{1/3}$, with V as in Eq.~\ref{eq:vol_defn}, we note that the horizon mass $M_{\rm H}$
is conserved, implying matter-domination-like expansion with $\rho_{\rm eff} \propto a^{-3}_{\rm FLRW}$.
There is no spin dependence to within numerical uncertainty. However, the time-dependence of the effective
Hubble parameter relative to an FLRW model shows a mix of matter and radiation contributions.
The bottom 3 lines in this figure imply that a purely matter dominated expansion $(a/a_0)^{-3}$ is not a good fit,
while the top 3 lines show the behavior is well-described by including radiative content with
a best fit function $0.19(a/a_0)^{-4} + 0.81(a/a_0)^{-3}$.

\section{Discussion and Conclusions}
\label{sec:conclusions}

In this paper, we have built a new kind of black-hole-lattice universe by
solving the constraint equations with a conformal-transverse-traceless decomposition with periodic boundary conditions.
A series of space-like hypersurfaces corresponding to expanding universes with spinning black holes
were identified. We found that the expansion features of those initial slices were very
close to an effective FLRW universe regardless of the initial parameters choices.

We then evolved the initial slices with a singularity-avoiding gauge choice,
finding no significant difference between the expansion in directions parallel and perpendicular to the spin.
The effective density described by the black hole mass evolved similarly to a matter-dominated universe,
ie. the mass of the black hole was conserved,
while the effective Hubble parameter only followed matter-dominated behavior at late times.
When quantifying the deviation of the expansion rate from FLRW-like behavior,
we found that averages taken over edges of the simulation coordinate box displayed large fluctuations, while volume averaged quantities showed much smaller deviations and approached FLRW asymptotically.
By fitting the spinning-black-hole-lattice universe to the FLRW universe, we were able 
to identify the effective mass that governs the expansion as the horizon mass of the black hole, 
rather than the irreducible mass.

In future work, we can track physical observables through the spacetime,
to better characterize the effects of highly non-linear non-stationary
perturbations on universes that, like are own, appear to be on-average
homogeneous and statistically isotropic on large scales.

It is noteworthy that the spacetimes that we have considered have a preferred direction, determined by the orientation of the spin of the single black hole in the fundamental domain.
We anticipate exploring more general initial conditions with multiple spinning black holes
and $0$ total angular momentum. 
Finally, although the spurious gravitational waves
introduced by initial gauge fixing (conformally flat and $\hat{A}^{TT}_{ij} = 0$) have been shown to not critically affect the late-time evolution in both previous work \cite{gr-qc/9710096, Burko:2005fa} and in our observation of $\tilde{A}_{ij} \tilde{A}^{ij}$, it may nevertheless be helpful to explore other
schemes, like conformal thin-sandwich (CTS) decomposition, to set more general initial conditions with reduced spurious gravitational wave content.

\section*{Acknowledgments}

We thank Tim Clifton, Mikołaj Korzyński, and Eloisa
Bentivegna for insightful discussions that helped shape this work.
This work benefited from the Sexten
Center for Astrophysics workshop on GR effects in cosmological
large-scale structure, and made use of the High Performance
Computing Resource in the Core Facility for Advanced Research Computing
at Case Western Reserve University.
JBM acknowledges support as a CITA National Fellow and from Perimeter Institute
for Theoretical Physics. Research at Perimeter Institute is supported by
the Government of Canada through the Department of Innovation, Science
and Economic Development Canada and by the Province of Ontario through
the Ministry of Research, Innovation and Science.  JTG is supported
by the National Science Foundation Grant No. PHY-1719652.
GDS and CT were supported in part by grant DE-SC0009946 from the US DOE.

\section*{\refname}
\bibliography{references}

\appendix

\section{Using puncture method to build initial data}
\label{sec:init}

In a black hole lattice, the metric near the black hole center will be
close to an isolated black hole, so
we can expect them to have similar divergence properties, i.e., 
$\Psi$ diverges
as $2M/r$ and $X^i$ diverges as $a x^i/r^3$ according
to the Bowen-York solution in Eqs. \ref{eq:XiBYinPB} and \ref{eq:PsiBYinPB} with
$J_i = (0, 0, a)$.

We therefore employ the puncture approach by defining
\begin{eqnarray}
  \label{eq:16}
  u & \equiv \Psi - \frac{M}{2r} (1-W(r)) \\ \nonumber
  X'^1 & \equiv X^1 - \frac{ya(1-W(r))}{r^3} \\ \nonumber
  X'^2 & \equiv X^2 + \frac{xa(1-W(r))}{r^3} \\ \nonumber
  X'^3 & \equiv X^3.
\end{eqnarray}
By switching from variables $(\Psi, X^i)$ to $(u, X'^i)$, we
expect to replace divergent variables with regular variables and implicitly
incorporate the divergence in the solution.

The constraint equations Eq. \ref{eq:simplifiedconstraints} can then be reduced to 
\begin{eqnarray}
  \label{E:17}
  &\nabla^2 u - \nabla^2\left(\frac{M}{2r} W(r)\right) + \frac{1}{8} \left( \tilde{L} X \right)_{ij} \left( \tilde{L} X \right)^{ij} \Psi^{-7} - \frac{1}{12} K^2 \Psi^5 = 0 \\ \nonumber
  &\nabla^2 X'^{1} + \frac13 \partial^1 \partial_j X'^j - \nabla^2 \left(\frac{ya }{r^3}W(r)\right) - \frac{2}{3} \Psi^6 \partial^1 K = 0 \\ \nonumber
    &\nabla^2 X'^{2} + \frac13 \partial^2 \partial_j X'^j + \nabla^2 \left(\frac{xa }{r^3}W(r)\right) - \frac{2}{3} \Psi^6 \partial^2 K = 0 \\ \nonumber
      &\nabla^2 X'^{3} + \frac13 \partial^3 \partial_j X'^j - \frac{2}{3} \Psi^6 \partial^3 K = 0,
\end{eqnarray}
which contain no divergent terms.
We solve for the variables $u$ and $X'^i$ using a standard multigrid method (more detail can be found in
\cite{Press_2003}), 
and the divergent initial data for $\Psi$ and $X^i$ can be restored from them.

\section{Numerical Convergence Details}
\label{sec:convergence}

We will show here detail of our convergence test. For three runs with different coarsest resolutions,
convergence rate is calculated as
\begin{eqnarray}
\label{eq:c}
 c \equiv \frac{|f_{N_c} - f_{N_m}|}{|f_{N_m} - f_{N_f}|},
\end{eqnarray}
where $f_{N_c}$, $f_{N_m}$ and $f_{N_f}$ are values calculated at resolutions $N_c$, $N_m$ and $N_f$, which are 
from coarsest to finest. Resolutions are chosen to be $64$, $96$ and $128$ respectively in our tests. 
As among all the simulations in this article the case R3 in Table \ref{t1} with spin parameter $a = 0.9$ exhibited the most instability, we will only focus on this case.

\begin{figure}[htbp]
  \centering
    \includegraphics[width=1.0\textwidth]{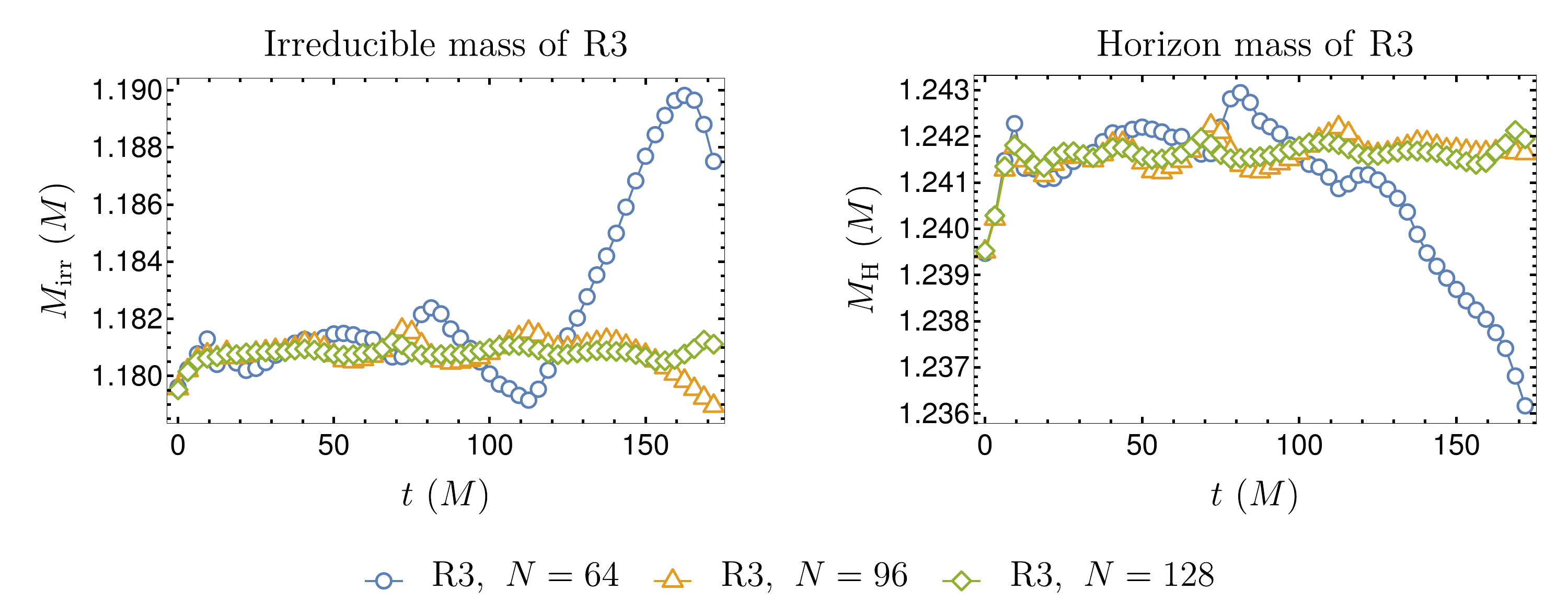}
  
  \caption{Behavior of irreducible mass (left) and horizon mass (right) with different coarsest resolutions in run R3 with $a = 0.9$.} 
  \label{mass_conv}
\end{figure}

We track the evolution history of both the irreducible mass and the horizon mass at
different resolutions in order to check for convergence. 
These are shown in Fig. \ref{mass_conv}, and both show small fluctuations that decrease
as numerical precision increases. To within this numerical error, the results we find are
consistent with the area theorem and with conservation of angular momentum.

\begin{figure}[htbp]
  \centering
    \includegraphics[width=1.0\textwidth]{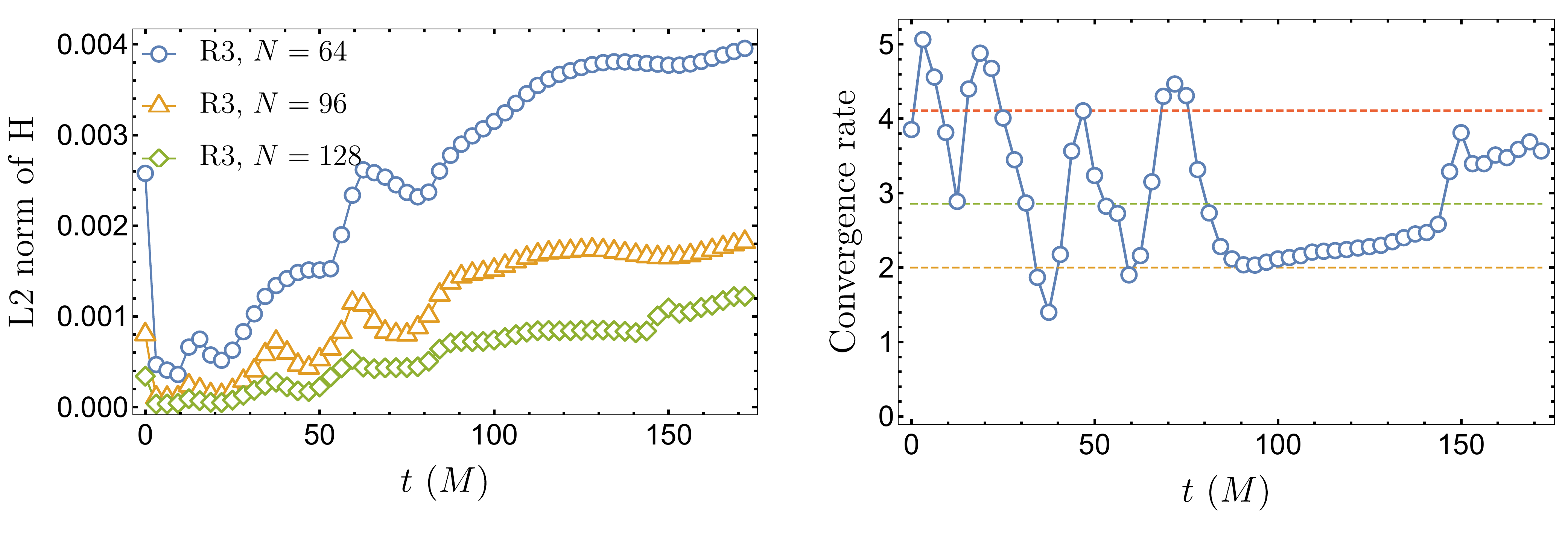}
  
  \caption{Behavior of the L2 norm of Hamiltonian constraint under different coarsest resolutions
    and the corresponding convergence rate. Dashed lines in the right plot indicates the 2nd, 3rd,
    and 4th order of convergence rate correspondingly from bottom to top.
    Second order of convergence can be achieved.} 
  \label{H_conv}
\end{figure}

The L2 norm of the Hamiltonian constraint at different resolutions, as well as the convergence rate, are
shown in Fig. \ref{H_conv}. Note that the L2 norm of the constraint violation is calculated
only outside of the black hole horizon.
Second order of convergence rate is achieved, as shown in the figure.
Note that during the evolution, AMR hierarchies are built even on the initial slice,
the interpolation operations used to build those levels reduce the convergence rate of initial violation from 7 (in the last column of Table \ref{t1}) to 4.
The mismatch between the 2nd order convergence and 4th order stencil mainly results from the truncation error introduced by coarse-fine interfaces.

\end{document}